# Goal-oriented Data Visualization with Software Project Control Centers

*Jens Heidrich, Jürgen Münch*

Fraunhofer Institute for Experimental Software Engineering

jens.heidrich@iese.fraunhofer.de, juergen.muench@iese.fraunhofer.de

## Zusammenfassung:

*Vielen Software-Entwicklungsorganisationen fehlt die notwendige Unterstützung, um quantitative Kontrolle über ihre Entwicklungsprozesse ausüben zu können, etwa, um die Prozessperformanz oder die Qualität erzeugter Produkte zu messen. Eine systematische Unterstützung zur Identifikation kritischer Projektsituationen und die Einplanung entsprechender Gegenmaßnahmen zur Erreichung der Projektziele fehlen üblicherweise. Ein Mittel, um Messen auf der Basis expliziter Modelle zu etablieren, ist die Entwicklung und Einführung so genannter Software-Projektleitstände zur systematischen Qualitätssicherung und Unterstützung des Projektmanagements. Ein Leitstand ist vergleichbar mit einem Kontrollzentrum im Bereich der mechanischen Produktion. Seine Aufgaben umschließen das Sammeln, Interpretieren und Visualisieren von Messdaten, um kontext-, zweck- und rollenbundene Informationen für alle beteiligten Interessengruppen während der Durchführung eines Software-Entwicklungsprojektes anzubieten (z.B. für Projektmanager, Qualitätssicherer oder Entwickler). Der vorliegende Artikel gibt einen Überblick über bestehende Leitstandlösungen und beschreibt eine konkrete Leitstandinstanz, die zielgerichtete Datenvisualisierung unterstützt (G-SPCC-Ansatz). Darüber hinaus werden Erfahrungen aus praktischen Anwendungen vorgestellt.*

## Schlüsselbegriffe

*Software-Projektleitstand, Dateninterpretation, Datenvisualisierung, zielgerichtetes Messen, Goal Question Metric Paradigm (GQM)*

## Abstract:

*Many software development organizations still lack support for obtaining intellectual control over their software development processes and for determining the performance of their processes and the quality of the produced products. Systematic support for detecting and reacting to critical project states in order to achieve planned goals is usually missing. One means to institutionalize measurement on the basis of explicit models is the development and establishment of a so-called Software Project Control Center (SPCC) for systematic quality assurance and management support. An SPCC is comparable to a control room, which is a well known term in the mechanical production domain. Its tasks include collecting, interpreting, and visualizing measurement data in order to provide context-, purpose-, and role-oriented information for all stakeholders (e.g., project managers, quality assurance manager, developers) during the execution of a software devel-*

*opment project. The article will present an overview of SPCC concepts, a concrete instantiation that supports goal-oriented data visualization (G-SPCC approach), and experiences from practical applications.*

*Keywords*

*Software Project Control Center (SPCC), Data Interpretation, Data Visualization, Goal-oriented Measurement, Goal Question Metric Paradigm (GQM)*

# 1    Introduction

The complexity of software development projects continues to increase. One major reason is the ever-increasing complexity of functional as well as non-functional software requirements (e.g., reliability or time constraints for safety critical systems). The more complex the requirements, the more people are usually involved in meeting them, which further increases the complexity of controlling and coordinating the project. This, in turn, makes it even harder to develop the system according to plan (i.e., matching time and budget constraints). Project control issues are very hard to handle. Many software development organizations still lack support for obtaining intellectual control over their software development projects and for determining the performance of their processes and the quality of the produced products. Systematic support for detecting and reacting to critical project states in order to achieve planned goals is usually missing [10].

One way to support effective control of software development projects is the use of basic engineering principles [2], [16], with particular attention to the monitoring and analysis of actual product and process states, the comparison of actual states with planned states, and the initiation of any necessary corrective actions during project execution. Effectively applying these principles requires experience-based project planning [13] and the use of explicitly defined models in order to plan a project. Furthermore, it requires the collection, interpretation, and presentation of measurement data according to a previously defined plan in order to provide stakeholders with up-to-date information about the project state. Moreover, it requires experience packaging after project completion so that future projects can profit from the experiences gained.

In the aeronautical domain, air traffic control systems are used to ensure the safe operation of commercial and private aircraft. Air traffic controllers use these systems to coordinate the safe and efficient movement of air traffic (e.g., to make certain that planes stay a safe distance apart or to minimize delays). The system collects and visualizes all critical data (e.g., the distance between two planes, the planned arrival and departure times) in order to support decisions by air traffic controllers. Software project control requires an analogous approach that is tailored to the specifics of the process being used (e.g., its non-deterministic, con-current, and distributed nature). A Software Project Control Center (SPCC) [10] is a control system for software development that collects all data relevant to project control, interprets and analyzes the data according to the project's control needs, visualizes the data for different project roles, and suggests corrective actions in the case of plan deviations. An SPCC could also support the packaging of data (e.g., as predictive models) for future use and contribute to an improvement cycle spanning a series of projects.

Controlling a project means ensuring the satisfaction of project objectives by monitoring and measuring progress regularly in order to identify variances from the plan during project execution, so that corrective action can be taken when necessary [11]. Planning is the basis for project control and defines expectations, which can be checked during project execution. Project control is driven by different role-oriented needs. We define control needs as a set of role-dependent requirements for obtaining project control. A project manager needs different kinds of data, data of different granularity, or different data visualizations than a quality assurance manager or a developer. For example, a manager is interested in an overview of the project effort in order to compare it to previously defined baselines, while a developer is interested in the effort spent on a certain activity.

In this article, we want to illustrate selected existing SPCC approaches (Section 2), and then focus on a concrete instantiation that supports goal-oriented data visualization, the so called G-SPCC approach (Section 3). Afterwards, we will present experiences gathered in applying this approach in practical courses at the University of Kaiserslautern and present some lessons learned (Section 4). Finally, we will give a summary and illustrate future research fields (Section 5).

## 2    Related Work

The following section presents an overview of selected SPCC approaches. In this, the scope has been defined as on-line data interpretation and visualization on the basis of past experience. A more detailed description of the strengths and weaknesses of the approaches presented can be found in [10].

Provence [8] is generally a framework for project management. It informs managers about state changes of processes and products and allows them to initiate dynamic re-planning steps. The basic architecture is focused on openness and adaptability with respect to integration into a specific software development organization.

Amadeus [15] is a metric-based analysis and feedback system embedded into the process-centered SDE Arcadia. The focus is on the integration of measurement as an active component within software development processes. Amadeus is based on a server/client architecture and uses active agents in order to interpret user de-fined scripts. It is possible to add new functionality by creating a new script or expanding the server's or client's tool kit. The data is processed with respect to the kinds of usage purposes implemented by appropriate techniques and methods, like classification tree or interconnectivity analysis.

Ginger2 [19] implements a computer-aided empirical software engineering (CAESE) environment and is focused on experimental aspects of software development, so-called in vitro studies. Ginger2 presents a framework for empirical studies, consisting of a life cycle model, data collection models, and a basic CAESE framework. It fulfills its tasks through a multitude of predefined data collection and analysis techniques, such as techniques to gather audio and video data of the experiment participants, data about mouse and window movements, and data about tool usages and program changes.

The SME (Software Management Environment) [6] is an automated management tool, which provides a predefined pool of techniques to observe, compare, predict, analyze, assess, plan, and control a software development project. It was developed within NASE/SEL [9]. The SME distinguishes three types of databases: one captures information about previous projects, a second one provides research results from studies of software development projects, and a third one includes management rules for guiding a project manager.

WebME (Web Measurement Environment) [18] is a successor of the SME approach and provides similar functionality. It enhances the capabilities of the SME in terms of supporting distributed development of software. It is built upon a three-layered, mediated architecture, which uses a web browser to access one of the WebME servers and data wrappers in order to collect data from distributed SDEs. It uses the Data Definition Language (DDL) for specifying scripts, which are used to determine data sources and data transformations.

The PPM (Process Performance Manager) [7] is a tool that is used to support the management of business processes. It is discussed here because of its open and portable architecture. The basic idea is to close the feedback gap between the specification and the execution of business processes. A number of source systems can be integrated via an XML interface. So-called key performance indicators (KPIs) can be visualized with the help of different views and filters, for instance, in order to identify deviations from nominal performance guidelines.

PAMPA [14] is a tool that is especially designed for data collection and visualization. It uses intelligent agents to reduce the overhead of data collection and analysis. For that, it provides a set of predefined objects with attributes and relationships among each other, which are instantiated for each project.

## 3 Goal-oriented Data Interpretation and Visualization

In this section, we want to illustrate a goal-oriented SPCC approach (called G-SPCC for short), which is a state-of-the-art framework for project control developed at the University of Kaiserslautern and the Fraunhofer Institute for Experimental Software Engineering (IESE) [3], [4], [5]. The aim of this approach is to interpret and present the collected data in a goal-oriented way in order to optimize a measurement program and effectively detect plan deviations. This section will present the high-level G-SPCC architecture, discuss the concepts of visualization catena, and finally describe the integration into the QIP [1] and the GQM paradigm [12], as part of a goal-oriented measurement framework.

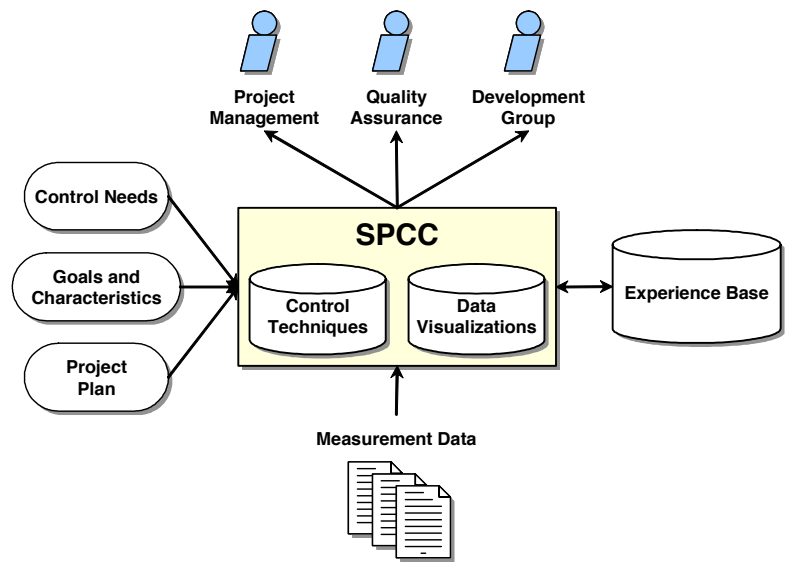

**Figure 1:** The high-level G-SPCC architecture.

Figure 1 gives an overview of the G-SPCC architecture. It shows that measurement data is collected during project performance and interpreted and visualized by the SPCC with respect to the goals and characteristics of the project as well as to project plan information and control needs. The core of an SPCC is a set of integrated project control techniques (for interpreting the data in the right way) and data visualization mechanisms (for presenting the interpreted data in accordance with the role interested in the data). The control techniques usually cover different purposes, such as monitoring project attributes, comparing attributes to baselines, or predicting the course of an attribute. The data visualization mechanisms provide role-specific insights into the process (e.g., insights suitable for project managers, quality assurance personnel, or the development group). The SPCC interpretation and visualization process is supported by an experience base in order to reflect data from previous projects and store experience gathered after project completion.

A so-called Visualization Catena (VC) [5] formally describes the relation between data collected, control techniques applied, and visualization mechanisms used. It basically consists of representations of the collected measurement data (so-called data entries), a selected and instantiated set of control techniques that interpret the collected data (so-called function instances), and a set of instantiated data visualization mechanisms that visualize the data in a role-oriented way (so-called view instances). An example can be found in Figure 2. The presented VC consists of a view suited for the project manager that basically displays effort data according to a defined process and checks for baseline deviations. Effort Tolerance Range Checking is applied to compare actual effort values with baseline values and to mark plan violations accordingly.

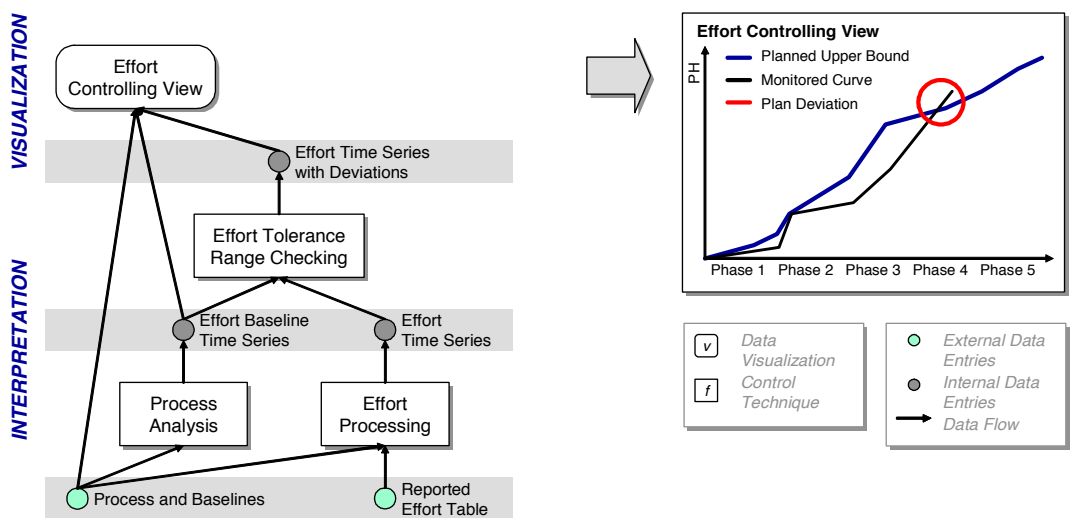

**Figure 2:** Sample Visualization Catena.

This approach can be used for conducting project control according to the following six steps (based on the QIP):

*Characterize Control Needs:* First, project stakeholder control needs are characterized in order to set up a measurement program that is able to provide a basis for satisfying all needs.

*Set Measurement Goals:* Then, measurement goals are defined and metrics are derived determining what kind of data to collect. The GQM paradigm is used to derive these metrics and create a set of data collection sheets that are assigned to certain process steps.

*Choose VC Components:* Next, a VC is defined to provide online feedback on the basis of the collected data; that is, control techniques and visualization mechanisms are selected from a corresponding repository and instantiated in the context of the project that has to be controlled.

*Execute the VC:* Once the VC is specified, a set of role-oriented views are generated for controlling the project. When measurement data are collected, the VC interprets and visualizes them accordingly, so that plan deviations can be detected.

*Analyze VC Results:* Once a deviation is detected, its root cause must be deter-mined and the control mechanisms have to be adapted accordingly (e.g., new parameters of the VC have to be set or a new baseline has to be chosen). This, does, for example, require the possibility to analyze data on different levels of abstraction in order to be able to trace causes for plan deviations.

*Package VC Results:* After project completion, the resulting VC may be used as a basis for defining control activities for future projects (e.g., selecting the right control techniques and data visualizations, choosing the right parameters for controlling the project).

## 4        G-SPCC Application

The following section describes the application of our approach in the context of practical courses at the University of Kaiserslautern (UKL). We will explain the setting of the studies and discuss some lessons learned from the application.

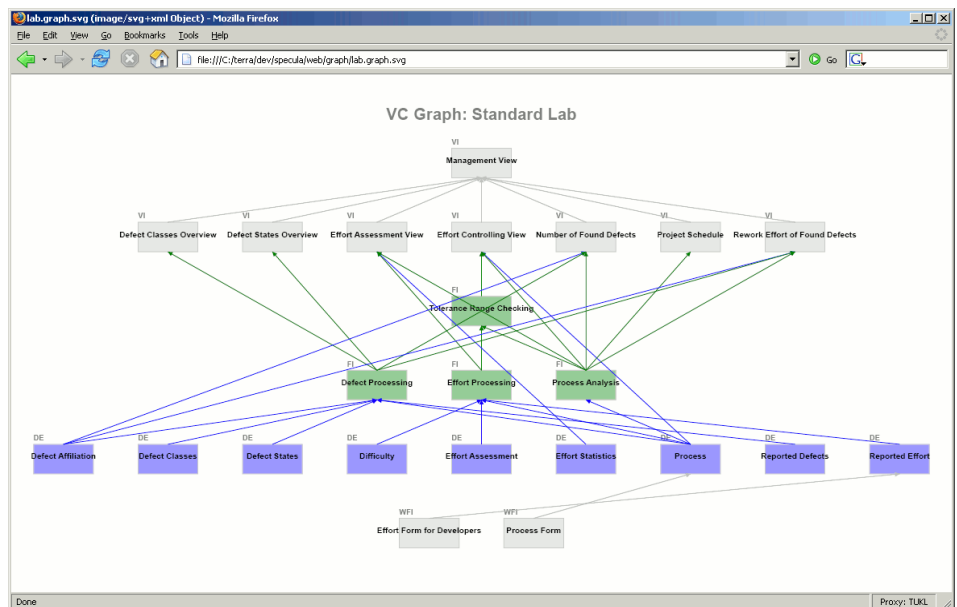

**Figure 3:**        VC used to control the practical course.

The described G-SPCC controlling approach was used in three different practical software engineering courses at the UKL, covering the areas of E-Commerce, building automation, and web-based applications, with 11, 14, and 43 students, respectively. Basically, we collected data concerning the invested effort of process steps and the efficiency of defect detection according to the following measurement goal specification (following the GQM paradigm): (1) Analyze the project effort for the purpose of baseline checking from the viewpoint of the project manager in the context of a practical course at UKL. (2) Analyze the found defects for the purpose of defect tracking from the viewpoint of the QA manager in the context of a practical course at UKL.

After goal definition, we derived metrics and created a set of data collection sheets that are assigned to certain process steps. The data collection process was supported by Glockenspiel [17], a light-weight process enactment machine, developed at the University of Kaiserslautern.

Next, we built up the VC to provide online feedback from the data collected. The G-SPCC prototype implementation, called Specula (Latin for watch tower), executed the VC accordingly; that is, extracting data from Glockenspiel, transforming them based on selected control techniques, and providing web-based visualizations. An overview of the VC used can be found in Figure 3. Once a deviation was detected, its reason had to be analyzed and the VC had to be adapted accordingly. The view used to control effort progression is presented in Figure 4.

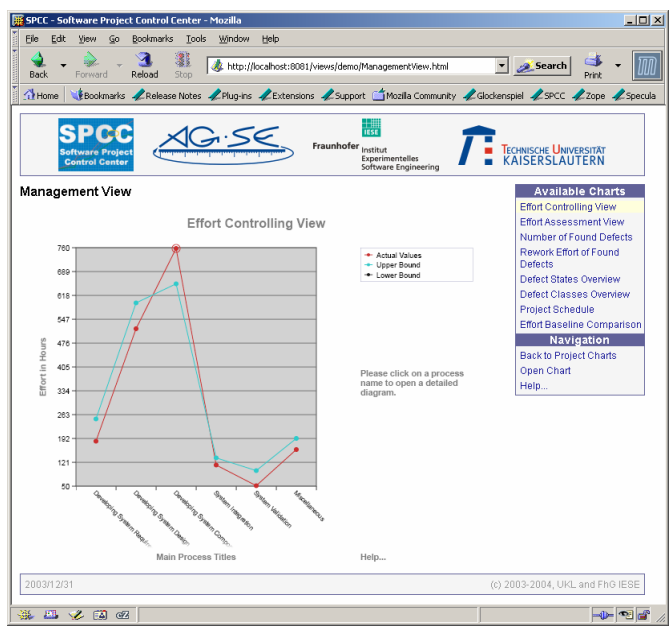

**Figure 4:** Project manager view of the applied VC.

Based on our experiences gathered through applying the G-SPCC approach, we could derive several lessons learned:

*Support for distributed development:* An SPCC has to provide different mechanisms in order to gather data from different development locations and to be adaptable to locally used tools for collecting the data (in our case, the Glockenspiel enactment environment).

*Data validity and consistency:* An SPCC has to provide mechanisms to test the plausibility of incoming data and a warning mechanism if invalid data is provided (e.g., by checking ranges or plausibility).

*Extensibility:* It is important to be able to add new control techniques and visualization mechanisms in order to better control the project. In our studies, e.g., we had to introduce new views in order to be able to have an assessment of the students with respect to the effort spent on the defined processes.

*Adaptability and variability:* The VC needs to be adapted (or parameterized) in order to match the current project characteristics. This includes, e.g., the tolerance range for baseline checking, as well as the graphical integration of the views produced (e.g., navigation structure and layout). For dynamic re-planning, it is important to be able to do this on-the-fly, meaning while project controlling activities are ongoing.

*Data abstraction and traceability:* If we want to detect the causes for a baseline deviation, it must be possible to have a more detailed view of the data. A more abstract view is suited for detecting such deviations, but not necessarily for finding the reasons behind them.

*Privacy capabilities:* If different groups use an SPCC, it is probably not desirable to give all users the same type of access to the data presented. In our case, students of different development groups were not supposed to be able to see each other's data, but the supervisor (as the manager of the overall project) was supposed to be able to do so.

*Support for reuse:* It is important to be able to reuse control mechanisms from previous projects in order to effectively set up project control. This includes control techniques (like simple Tolerance Range Checking in the courses or more advanced techniques like Value-based approaches), role-oriented data visualizations, and baselines from previous projects.

## 5 Conclusion

This article presented the basic concept of an SPCC as a means for establishing project control. We illustrated existing approaches and presented a goal-oriented way to establish project control by formalizing the way measurement data are interpreted and visualized according to a previously defined measurement goal. Existing approaches offer mostly partial solutions. Especially goal-oriented usages based on a flexible set of techniques and methods are not comprehensively supported [10]. The expected benefits of the G-SPCC approach include: (1) improvement of quality assurance and project control by providing a set of custom-made views of measurement data, (2) support of project management through early detection of plan deviations and proactive intervention, (3) support of distributed software development by establishing a single point of control, (4) enhanced understanding of software processes, and improvement of these processes, via measurement-based feedback, and (5) preventing information overload through custom-made views with different levels of abstraction.

The G-SPCC approach is part of ongoing research. An important research issue in this context is the development of a schema for adaptable control techniques and methods, which effectively allows for purpose-driven usage of an SPCC in vary-ing application contexts. Another research issue is the elicitation of information needs for the roles involved and the development of mechanisms for generating adequate role-oriented visualizations of the project data. Another important research issue is support of change management. When the goals or characteristics of a project change, the real processes react accordingly. Consequently, the control mechanisms, which should always reflect the real world situation, must be updated. This requires flexible mechanisms that allow for reacting to process variations. One long-term goal of engineering-style software development is to control and forecast the impact of process changes and adjustments on the quality of the software artifacts produced and on other important project goals. An SPCC can be seen as a valuable contribution towards reaching this goal.

**Acknowledgements**

We would like to thank Sonnhild Namingha from Fraunhofer IESE for reviewing the article. This work was partly funded by the Deutsche Forschungsgemeinschaft as part of the Special Research Project SFB 501 "Development of Large Systems with Generic Methods". In addition, the work was motivated by the project "Soft-Pit" funded by the Federal Ministry of Education and Research (BMBF) in the context of the research initiative "Software Engineering 2006".